\documentclass{article}

\usepackage{amsmath,amsthm}
\newtheorem{lemma}{Lemma}
\newtheorem{theorem}{Theorem}
\newtheorem{definition}{Definition}
\usepackage{graphicx}
\usepackage[ruled, lined, linesnumbered, boxed]{algorithm2e}
\usepackage{amssymb}
\usepackage{mathrsfs}
\usepackage{enumerate}
\usepackage{authblk}

\usepackage{geometry}
\usepackage{setspace}
\begin{document}

\title{Quantum Eigensolver for Non-Normal Matrices via Ground State Energy Estimation}

\author[1,2]{Honghong Lin}
\affil[1]{Institute of Mathematics, Academy of Mathematics and Systems Science, Chinese Academy of Sciences, Beijing 100190, China}
\affil[2]{School of Mathematical Sciences, University of Chinese Academy of Sciences, Beijing 100049, China}
\author[1,3]{Yun Shang\footnote{shangyun@amss.ac.cn}}
\affil[3]{State Key Laboratory of Mathematical Science, Academy of Mathematics and Systems Science, Chinese Academy of Sciences, Beijing 100190, China}

\maketitle

\begin{abstract}
Non-normal matrix eigenvalue problems arise in a wide variety of fields ranging from non-Hermitian physics to dynamical systems. State-of-the-art algorithms suffer from a prohibitive scaling with the condition number and target precision. In this work, we introduce a quantum eigensolver that significantly reduces the computational cost. We develop an estimation approach via an iteratively refined sampling scheme and reduce the error evaluation to ground state energy estimation, for which near-optimal algorithms exist. Our algorithm achieves a polynomial speedup by improving the gate count scaling in both the condition number and the target precision. We further extend our method to approximate different types of spectral gaps. An additional advantage of our work is its ability to directly prepare the eigenstate of the estimated eigenvalue using ground state preparation. We also discuss its applications in analyzing the stability of dynamical systems and in solving polynomial root-finding problems. Numerical simulations are presented to validate the performance of our algorithms.
\end{abstract}

\section{Introduction}


Quantum phase estimation \cite{nielsen2010quantum}, one of the most important quantum primitives, is an algorithm for estimating the eigenphases of a unitary matrix, which is mathematically equivalent to eigenvalue estimation for Hermitian matrices \cite{PRXQuantum.4.020331}. Solving eigenvalues for non-unitary or non-Hermitian matrices is a natural extension of quantum phase estimation. It has promising applications in, for example, the field of non-Hermitian physics. Specifically, real physical systems are to some extent coupled to their environment and the quantum dynamics of an open system is, in general, formulated by an equation of motion for its density matrix, a quantum master equation:
\begin{equation}
\frac{d}{dt}\rho(t) = \mathcal{L}(t)\rho(t).
\end{equation}
Here $\mathcal{L}(t)$ is the non-Hermitian Liouvillian superoperator \cite{breuer2002theory}. A fundamental obstacle in the analysis of these systems is the need to accurately resolve their full eigenvalue spectra. One of the fascinating phenomena in a non-Hermitian system is the non-Hermitian skin effect (NHSE) \cite{zhang2022review} where eigenstates exhibit localized behaviors, drastically different from the extended Bloch waves in Hermitian systems. Ref. \cite{Yang2025Commun} demonstrates that the NHSE can profoundly alter how particles interact and localize. Another fundamental concept in non-Hermitian physics is PT symmetry breaking \cite{vvrx-mljg}. In the PT-symmetric phase, the system exhibits a purely real energy spectrum, despite the Hamiltonian being non-Hermitian. However, when the system parameter exceeds some critical value, the PT symmetry breaks spontaneously, and the energy spectrum becomes complex. This extreme sensitivity to external perturbations makes PT-symmetric systems ideal for developing next-generation sensors with unprecedented resolution \cite{PhysRevLett.125.240506}.

The operation number required for classical eigenvalue estimation algorithms scales polynomially with respect to the matrix size, hence classical computers are limited in their ability to accurately estimate eigenvalues for exponentially large systems. Quantum computers, in contrast, give an exponential compression of the amount of memory space required to store a matrix, and therefore have potential to provide a significant speedup in solving such problems. 

Eigenvalue problems for non-normal matrices are hard due to two main reasons. One is that the eigenvalues can be complex numbers, whereas the measurement output of a quantum circuit is real. The other is the lack of a complete eigenstate basis if the matrix is not diagonalizable, while eigen-decomposition is the key to algorithmic analysis of many quantum eigenvalue algorithms for unitary or Hermitian matrices \cite{ni2023low,ding2023simultaneous}.

Although there have been numerous studies, existing algorithms either make strong assumptions or exhibit poor performance, and an efficient eigensolver for general non-normal matrices remains an open problem. In previous work, Shao \cite{shao2022computing} solved the case of diagonalizable matrices with real eigenvalues with success probability close to 1. For complex eigenvalues, the algorithm relies on a particular form of the initial state, the preparation of which is not clear. Low \& Su \cite{low2024quantum} took a step further and considered matrices with only real eigenvalues. Their approach to estimate the argument of the largest-modulus complex eigenvalue also relies on an oracle that approximately generate the associated eigenstate. Zhang et al. \cite{3n8f-k8pl} gave the first general algorithm for non-Hermitian matrices. They developed a singular value threshold subroutine, which provides a mechanism for eigenvalue detection. A key limitation, however, is the undesirable scaling behavior of this subroutine's gate complexity with respect to the algorithmic parameters.

\subsection{Contributions and Methodology}

In this paper, we present a quantum eigensolver for non-normal matrices that beats all previous work. The goal is quite general: given a non-normal matrix $A$, output an eigenvalue of $A$ with prescribed precision and success probability. Our strategy is to sample points on the unit disk and then assess their distance to the eigenvalues of $A$. The analytical tool we apply is a bound on the distance between a sampled point and the closest eigenvalue, which is related to a quantity $\sigma_0(\mu)$, the smallest singular value of $A-\mu I$. Hence, we are able to reduce the problem to ground state energy estimation via Hermitianizing $A-\mu I$. With the information of the estimation error bound, we can iteratively update the guessing region and obtain an estimate after sufficient steps. To demonstrate the flexibility of our method, we develop algorithms for approximating different types of spectral gaps, namely the point gap and line gap problem, which is the distance between a reference point or line and the closest eigenvalue. We also mention a scheme for estimating multiple eigenvalues.

To deal with non-normal matrices, we assume access to the block encoding input model, which encodes the target
matrix (rescaled) in a larger unitary matrix with the help of ancilla qubits. Then the Hermitianization can be realized using the powerful Quantum Singular Value Transformation (QSVT) technique. 

The main quantum subroutine of our algorithm is the ground state energy estimation, which is a hard problem even for quantum computer: deciding whether the ground energy of a generic local hamiltonian is smaller than $a$ or greater than $b$ for some $a<b$ is QMA-complete \cite{kitaev2002classical,kempe2006complexity}. Therefore, the near-optimal ground state preparation and energy estimation algorithm from Ref. \cite{lin2020near} relies on some assumptions. First, they assume access to an initial state which can be efficiently prepared by an oracle and has overlap with the ground state lower bounded by $\gamma$. Such an oracle can be realized by state preparation using variational quantum eigensolvers \cite{mcclean2016theory,peruzzo2014variational} or the boosting technique \cite{wang2022state}. They also assume a lower bound $\Delta$ on the spectral gap, which also can be estimated \cite{russo2021evaluating}. We inherit these assumptions in this work.

The computational cost is measured by a full characterization of the gate cost, including the gate complexity of the initial state preparation and the block encoding oracle of $A$. We also analyze the number of ancilla qubits required. We summarize the complexity improvements compared with the state-of-the-art algorithm \cite{3n8f-k8pl} in Table \ref{comparison} in terms of the query complexity of the block encoding oracle $U_A$. For the case of $m=1$, both work for eigenvalue estimation require $O(1/\epsilon)$ query complexity, which matches the Heisenberg limit for quantum phase estimation---also corresponding to $m=1$. For $m>1$, however, our algorithms achieve an improvement by a factor of $\kappa\epsilon^{m-1}$. This improvement is substantial, especially when high precision estimation is required. It is the reduction to ground energy estimation that gives this overhead saving. Also, our algorithms will directly benefit from future advances in ground state energy estimation.

Our algorithm establishes a superior paradigm for quantum non-normal eigensolver, characterized by:\\
{\bf Provable Speedup}: By reducing the problem to near-optimal ground state energy estimation, our algorithms bypass the need for less efficient subroutines and achieve a polynomial speedup in query complexity over the best known algorithms \cite{3n8f-k8pl}.\\ 
{\bf Direct Eigenstate Preparation}: With the knowledge of an eigenvalue, we are able to directly prepare the associated eigenstate through ground state preparation, a feature absent from existing techniques.\\
{\bf Built-in Flexibility}: One can design estimation algorithms for eigenvalues locating in different region or for different types of eigenvalue problems by modifying the sampling scheme.

\begin{table}
\centering
\caption{A comparison of the performance between this work and \cite{3n8f-k8pl} in terms of the query complexity of the block encoding oracle $U_A$.}\label{comparison}
\begin{tabular}{c|c|c|c}
\hline
& eigenvalue estimation & point gap estimation & line gap estimation \\
\hline
This work & $\widetilde{O}(\kappa^2\gamma^{-1}\epsilon^{-2m+1})$ & $\widetilde{O}(\kappa^2\gamma^{-1}\epsilon^{-2m})$ & $\widetilde{O}(\kappa^2\gamma^{-1}\epsilon^{-2m})$\\
Zhang et al. \cite{zhang2024exponentialquantumadvantagespractical} & $\widetilde{O}(\kappa^3\gamma^{-1}\epsilon^{-3m+2})$ & $\widetilde{O}(\kappa^3\gamma^{-1}\epsilon^{-3m+1})$ & $\widetilde{O}(\kappa^3\gamma^{-1}\epsilon^{-3m+1})$ \\
\hline
\end{tabular}
\end{table}

The rest of the paper is organized as follows. We present our basic algorithm for eigenvalue estimation in Section \ref{alg} and detail the implementation of the algorithm in Section \ref{implement}. In Section \ref{main_result}, we extend the basic algorithm to various eigenvalue problems and state our main complexity result. We mention two applications in Section \ref{app} and discuss some relevant issues in Section \ref{dis}.

\section{Algorithm}\label{alg}

Any matrix $A$ can be reduced by similarity transformation to a block diagonal matrix $J = P^{-1}AP$ consisting of $p$ diagonal blocks, each associated with a distinct eigenvalue $\lambda_j$ (referred to as Jordan matrices $J(\lambda_j)$). The dimension of $J(\lambda_j)$ is equal to the algebraic multiplicity of $\lambda_j$. Each Jordan matrix has itself a block diagonal structure consisting of $g_j$ subblocks (called Jordan blocks), where $g_j$ is the geometric multiplicity of the eigenvalue $\lambda_j$. Each of the Jordan block has the following structure
\begin{equation}
\begin{pmatrix}
\lambda_j & 1 & & \\
& \ddots & \ddots & \\
& & \lambda_j & 1 \\
& & & \lambda_j
\end{pmatrix}.
\end{equation}
The condition number $\|P\|\|P^{-1}\|$ varies with the scaling of the columns of $\|P\|$, hence we define the Jordan condition number to be $\inf_P \|P\|\|P^{-1}\|$ as in \cite{low2024quantum} and assume the knowledge of an upper bound $\kappa$.

An eigenvalue is said to be semi-simple if its algebraic multiplicity is equal to its geometric multiplicity, and
a matrix is diagonalizable if and only if all its eigenvalues are semi-simple. An eigenvalue whose algebraic multiplicity is larger than its geometric multiplicity is called defective, and a defective matrix is one that has at least one defective eigenvalue.

A simple observation is that if $\mu$ is an eigenvalue of $A$, then $A-\mu I$ is singular and the smallest singular value of $A-\mu I$ is zero. Let $\sigma_0(\mu)$ denote the smallest singular value of $A-\mu I$. Our eigenvalue estimation algorithm is based on the following lemma from Ref.\cite{3n8f-k8pl} with modifications.

\begin{lemma}\label{lemma}
Let $A = PJP^{-1}$ be the Jordan decomposition of $A$. Suppose that $\|A\| \le 1$ and the Jordan condition number $\|P\|\|P^{-1}\| \le \kappa$. Let $m_{\rm max}$ denote the block size of the largest Jordan block. For any $\mu$, $|\mu| \le 1$, $\sigma_0(\mu)$ is the smallest singular value of $A-\mu I$. Then the minimum distance of $\mu$ to the eigenvalues $\lambda_j$ of $A$ can be bounded as
\begin{equation}\label{diag}
\sigma_0(\mu) \le \min\limits_j|\mu-\lambda_j| \le \kappa\sigma_0(\mu)
\end{equation}
if $A$ is diagonalizable, and
\begin{equation}\label{defective}
\sigma_0(\mu) \le \min\limits_j|\mu-\lambda_j| \le 3(\kappa\sigma_0(\mu))^{1/m}
\end{equation}
for some $m \le m_{\rm max}$ if $A$ is defective.
\end{lemma}
\proof See Appendix \ref{proof_lemma_1}.

Note that for normal matrices, $\kappa=1$ and $m=1$, and Eq.\eqref{diag} becomes the trivial identity
\begin{equation}
\sigma_0(\mu) = \min\limits_j |\mu-\lambda_j|,
\end{equation}
which can be directly obtained from singular value decomposition. Equation \eqref{diag} for diagonalizable matrices is a rephrasing of the Bauer-Fike Theorem \cite[Theorem 2.3]{trefethen2020spectra}. Hence, Lemma \ref{lemma} generalizes the Bauer-Fike Theorem to defective matrices. We detail this in Appendix \ref{pseudospectra}.

\begin{figure*}[t]
\centering
\includegraphics[scale=0.45]{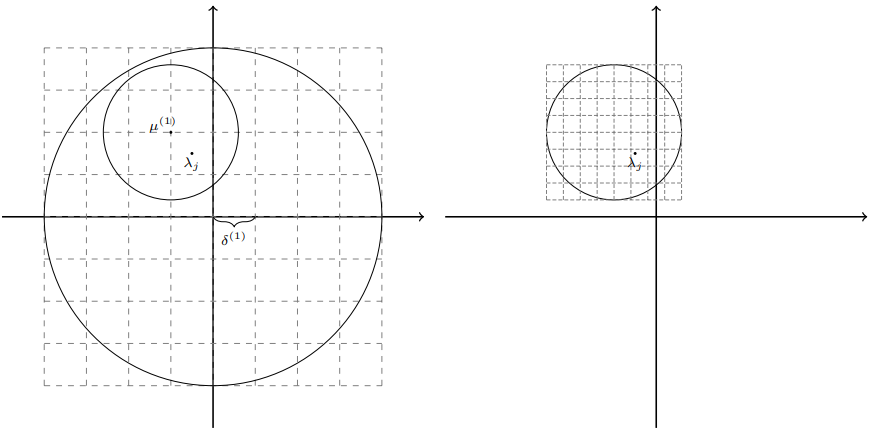}
\caption{\label{SearchRegion}Illustration of the first and second steps of Algorithm \ref{QEE}. Grid points (left) with spacing $\delta^{(1)}$} are candidates for $\mu^{(1)}$. Grid points (right) with spacing $\delta^{(2)}$ are candidates for $\mu^{(2)}$.
\end{figure*}

Our idea of estimating the eigenvalue is to use the upper bound of equation \eqref{diag} or \eqref{defective} to update the sampling region, starting with the unit disk. As depicted in Fig. \ref{SearchRegion}, at the first step, the grid points with spacing $\delta^{(1)}$ covering the unit disk are the candidates for $\mu^{(1)}$. If $\sigma_0(\mu^{(1)}) \le \delta^{(1)}$ for some $\mu^{(1)}$, we then know that $\min_j|\mu^{(1)}-\lambda_j| \le 3(\kappa\sigma_0(\mu^{(1)}))^{1/m} \le 3(\kappa\delta^{(1)})^{1/m}$. Hence, we can update the sampling region from the unit disk to the circular area centered at $\mu^{(1)}$ with radius $3(\kappa\delta^{(1)})^{1/m}$. At step $j \ge 2$, we choose grid points with spacing $\delta^{(j)}$. The parameters are chosen so that the sampling radii are reduced by a factor of 2 at each step. Therefore, after $O(\log(1/\epsilon))$ steps, we shrink the sampling region to size of $O(\epsilon)$ and obtain an $\epsilon$-estimate of an eigenvalue. See Algorithm \ref{QEE}. This procedure can also be interpreted from the perspective of pseudospectra. We elaborate on this in Appendix \ref{pseudospectra}.

\begin{algorithm}[t]
	\caption{Quantum Eigenvalue Estimation Algorithm}
	\label{QEE}
    \SetAlgoLined
    
    \KwIn{Unitary $U_A$ that block encodes $A$, target precision $\epsilon$.}
    \KwOut{Estimate of an eigenvalue of $A$ to precision $\epsilon$.}
    
    center = 0;  //initial sampling center \\
    $R^{(1)} = 1$; //initial sampling radius \\
    \For{$l = 1:\lceil\log(1/\epsilon)\rceil$}{
        $\delta^{(l)} = 1/(\kappa(3\cdot2^l)^m)$; //grid spacing \\
        $s = \lceil R^{(l)}/\delta^{(l)} \rceil$; \\
        BreakInner = false; \\
        \For{$j = -s:s$}{
            \For{$k = -s:s$}{
                $\mu^{(l)}_{j,k}$ = center + $j \cdot \delta^{(l)}+(k \cdot \delta^{(l)})i$; //$i$ is the imaginary unit\\
                construct the block encoding of $(A-\mu^{(l)}_{j,k} I)/\alpha_\mu$ with normalization factor $\alpha_\mu = 1+|\mu^{(l)}_{j,k}|$; \\
                block encode $H_\mu = \sqrt{\big((A-\mu^{(l)}_{j,k} I)^\dag(A-\mu^{(l)}_{j,k} I)/\alpha_\mu^2+\nu I\big)/(1+\nu)}$; \\
                call ground state energy estimation algorithm to $H_\mu$, return $E_0(\mu^{(l)}_{j,k})$; \\
                $\sigma_0(\mu^{(l)}_{j,k}) = \sqrt{E_0^2(\mu^{(l)}_{j,k})-\nu}$; \\
                \If{$\sigma_0(\mu^{(l)}_{j,k}) \le \delta^{(l)}$}{
                    center = $\mu^{(l)}_{j,k}$; \\
                    $R^{(l+1)} = 3(\kappa\sigma_0(\mu^{(l)}_{j,k}))^{1/m}$; \\
                    BreakInner = true; \\
                }
                \If {$\rm BreakInner$}{
                break; \\
                }
            }
            \If {$\rm BreakInner$}{
                break; \\
            }
        }
        
    }
\end{algorithm}

\section{Implementation}\label{implement}

Assume the input matrix $A \in \mathbb{C}^{N \times N}$ where $N=2^n$ is given in the block-encoding form. $U_A$ is called an $(\alpha,a,\epsilon)$-block-encoding \cite{gilyen2019quantum} of a $n$-qubit matrix $A$ using $a$ ancilla qubits if
\begin{equation}
\| A - \alpha(\langle0^a| \otimes I_n)U_A(|0^a\rangle \otimes I_n) \| \le \epsilon,
\end{equation}
where $\alpha$ is referred to as the normalization factor. Many matrices of practical interests can be efficiently block-encoded, such as sparse-access matrices \cite{gilyen2019quantum}. For simplicity we assume that the block encoding can be constructed without error in this work, and that $\|A\|\le1$, which implies that the eigenvalues lie within the unit circle of the complex plane. For some $\mu \in \mathbb{C}$, one can block encode the shifted matrix $A-\mu I$ with normalization factor $\alpha_\mu = 1+|\mu|$ from the block encoding of $A$, as is done in Ref. \cite[Lemma 7]{zhang2024exponentialquantumadvantagespractical}. The resulting unitary is referred to as $U_\mu$.

One key step of Algorithm \ref{QEE} is the evaluation of the smallest singular value of $A-\mu I$. Suppose that the singular values of $A-\mu I$ in ascending order are $\sigma_0, \sigma_1,...,\sigma_{N-1}$, then $H_{\mu} = \sqrt{(A-\mu I)^{\dagger}(A-\mu I)}$ is a Hermitian matrix with eigenvalues $\sigma_0, \sigma_1,...,\sigma_{N-1}$. That is, the smallest singular value of $A-\mu I$ equals the ground state energy of $H_{\mu}$. Hence we can evaluate $\sigma_0(\mu)$ by applying the ground state energy estimation algorithm \cite{lin2020near} to the Hermitian Hamiltonian $H_{\mu}$.

The Hamiltonian $H_\mu$ is a matrix function of the Hermitian matrix $G_\mu = (A-\mu I)^\dag(A-\mu I)$, and $G_\mu$ can be block encoded with normalization factor $\alpha_\mu^2$ through matrix multiplication of $U_\mu$ \cite[Lemma 30]{gilyen2019quantum}. One can construct $H_\mu$ by approximating the square function. Since the square function is not analytic on $[0,1]$ but on $[\eta,1]$ for some $\eta \in (0,1)$, we shift $G_\mu/\alpha_\mu^2$ by $\nu I$ for some $\nu > 0$, whose block encoding is demonstrated in Fig. \ref{BE_circuit} in Appendix \ref{proof_lemma_2}. $H_\mu$ is now redefined as
\begin{equation}\label{H_mod}
H_\mu = \sqrt{\frac{(A-\mu I)^\dag(A-\mu I)/\alpha_\mu^2+\nu I}{1+\nu}}.
\end{equation}
Then we can approximate the square function on $[\nu/(1+\nu),1]$ by some polynomial. Quantum signal processing (QSP) technique \cite{gilyen2019quantum} enables us to perform eigenvalue transformation of a Hermitian matrix for polynomials of arbitrary parity.

\begin{lemma}\label{H_mu}
There is a $(1,2a+5,\epsilon)$-block-encoding $U_H$ of $H_\mu$ using $O(\log(1/\epsilon))$ queries of $U_A$ and $O((2a+4)\log(1/\epsilon))$ primitive gates.
\end{lemma}
\proof See Appendix \ref{proof_lemma_2}.

\section{Error analysis and gate complexity}\label{main_result}

As mentioned in Introduction Section, we follow the assumptions for the ground state algorithms from \cite{lin2020near}. Let $O^{\mu}_{\rm init}$ be the initial state preparation oracle such that the initial state prepared has overlap with the ground state of $H_{\mu}$ lower bounded by $\gamma$. Also, we assume the spectral gap of $H_\mu$ is lower bounded by $\Delta$ for all sampled values of $\mu$.

The following theorems are the main results of our work. Here, $C_A$ denotes the gate cost for implementing the oracle $U_A$ and $C_{\rm init}$ is the maximum gate cost of $O^{\mu}_{\rm init}$ over all sampled values of $\mu$. We hide logarithmical dependence on the parameters in the $\widetilde{O}$ notation.

\begin{theorem}[Quantum eigenvalue estimation]\label{QEE_thm}
Suppose that a matrix $A \in \mathbb{C}^{N \times N}$ is given through a $(1,a,0)$-block-encoding $U_A$. Then there is an algorithm that, with probability at least $1-p_{\rm fail}$, outputs an $\epsilon$-estimate of an eigenvalue of $A$ using $O(a+\log(1/\gamma))$ ancilla qubits with gate count
\begin{equation}
\widetilde{O}\Big(\kappa^2\gamma^{-1}\epsilon^{-2m+2}\big(C_{\rm init}+\epsilon^{-1}(a+C_A)\big)\Big).
\end{equation}
\end{theorem}

\begin{proof}
At the $l$-th sampling step, we set the grid spacing $\delta^{(l)} = 1/(\kappa(3\cdot2^l)^m)$. If $\sigma_0(\mu^{(l)}_{j,k}) \le \delta^{(l)}$, then
\begin{equation}
\begin{aligned}
\min\limits_j |\mu^{(l)}_{j,k}-\lambda_j| 
& \le 3(\kappa\sigma_0(\mu^{(l)}_{j,k}))^{1/m} \\
& \le 3(\kappa\delta^{(l)}_{j,k})^{1/m} \\
& \le \frac{1}{2^l}.
\end{aligned}
\end{equation}
Hence, at step $L = \lceil\log(1/\epsilon)\rceil$, for $\mu^{(L)}_{j,k} = \mu^{(L)}_{j_0,k_0}$ that satisfies $\sigma_0(\mu^{(L)}_{j_0,k_0}) \le \delta^{(L)}$, we have
\begin{equation}
\min\limits_j |\mu^{(L)}_{j_0,k_0}-\lambda_j| \le \frac{1}{\epsilon}.
\end{equation}
That is, $\mu^{(L)}_{j_0,k_0}$ is an $\epsilon$-estimate of an eigenvalue of $A$.

The number of sampling points at step $l$ is upper bounded by
\begin{equation}
N^{(l)} = \left\lceil \frac{2R^{(l)}}{\delta^{(l)}} \right\rceil^2 = O(\kappa^22^{2(m-1)l}).
\end{equation}
Then the total number of sampling $N_s$ is upper bounded by
\begin{equation}
\sum_{l=1}^{L} N^{(l)} = \kappa^2 \sum_{l=1}^{L} O(2^{2(m-1)l}) = O(\kappa^2\epsilon^{-2(m-1)}).
\end{equation}
The query complexity is $N_s$ times the query number required by the ground state energy estimation. The near-optimal algorithm from Ref. \cite{lin2020near} estimates ground energy through binary amplitude estimation and is based on a polynomial approximating the sign function. Due to the robustness of quantum eigenvalue transformation \cite[Lemma 22]{gilyen2019quantum}, the block encoding error in Lemma \ref{H_mu} propagates to the error in the polynomial transform while introducing a factor of $n$, which we hide in the big-$O$ notation for simplicity. To achieve precision $\epsilon$ and success probability $1-\theta$, the algorithm uses 
\begin{enumerate}[1)]
  \item $O(a+\log(1/\gamma))$ ancilla qubits,
  \item $N_I'$ times initial state preparation, where $N_I' = O\big(\frac{1}{\gamma}\log\big(\frac{1}{\epsilon}\big)\log\big(\frac{1}{\theta}\log\big(\frac{1}{\epsilon}\big)\big)\big )$,
  \item $O\big(\frac{1}{\epsilon}\log\big(\frac{1}{\gamma}\big)N_I'\big)$ queries of $U_H$, and
  \item $O\big(\frac{a}{\epsilon}\log\big(\frac{1}{\gamma}\big)N_I'\big)$ other one- and two-qubit gates.
\end{enumerate}
It is sufficient to take $\theta = p_{\rm fail}/N_s$ to ensure success probability $1-p_{\rm fail}$ of our algorithm, hence the number of queries to the initial state preparation oracles is
\begin{equation}
\begin{aligned}
  N_I & = N_s \cdot N_I' \\
  & = O\left(\frac{\kappa^2}{\epsilon^{2(m-1)}}\right) \cdot  O\left(\frac{1}{\gamma}\log\left(\frac{1}{\epsilon}\right)\log\left(\frac{N_s\log(1/\epsilon)}{p_{\rm fail}}\right)\right) \\
& = O\left(\frac{\kappa^2}{\gamma\epsilon^{2(m-1)}}\log\left(\frac{1}{\epsilon}\right)\log\left(\frac{\kappa^2}{p_{\rm fail}} \frac{\log(1/\epsilon)}{\epsilon^{2(m-1)}}\right)\right).
\end{aligned}
\end{equation}
Note that one query of $H_\mu$ uses $O(\log(1/\epsilon))$ queries of $U_A$ and $O((2a+4)\log(1/\epsilon))$ primitive gates. Therefore, the total query complexity of $U_A$ is
\begin{equation}
\begin{aligned}
&N_s \cdot O\left(\frac{1}{\epsilon}\log\left(\frac{1}{\gamma}\right)N_I'\right) \cdot O\left(\log\left(\frac{1}{\epsilon}\right)\right) \\= &O\left(\frac{1}{\epsilon}\log\left(\frac{1}{\gamma}\right)\log\left(\frac{1}{\epsilon}\right)N_I\right),
\end{aligned}
\end{equation}
and the number of one- and two-qubit gates is
\begin{equation}
\begin{aligned}
&N_s \cdot \bigg[ O\left(\frac{a}{\epsilon}\log\left(\frac{1}{\gamma}\right)N_I'\right) + O\left(\frac{1}{\epsilon}\log\left(\frac{1}{\gamma}\right)N_I'\right) \cdot O\left( (2a+4)\log\left(\frac{1}{\epsilon}\right) \right)  \bigg] \\ = &O\left(\frac{a}{\epsilon}\log\left(\frac{1}{\gamma}\right)\log\left(\frac{1}{\epsilon}\right)N_I\right).
\end{aligned}
\end{equation}
\end{proof}

A special case of Theorem \ref{QEE} is that the eigenvalues are all real, and this does occur in real quantum systems \cite{PhysRevLett.80.5243}. When matrix $A$ has only real eigenvalues, the following theorem shows the query complexity can be substantially reduced.

\begin{theorem}[Real eigenvalues]\label{QEE_real}
Suppose that matrix $A \in \mathbb{C}^{N \times N}$ is given through a $(1,a,0)$-block-encoding $U_A$ and has only real eigenvalues. Then there is an algorithm that, with probability at least $1-p_{\rm fail}$ outputs an $\epsilon$-estimate to an eigenvalue of $A$ using $O(a+\log(1/\gamma))$ ancilla qubits with gate count
\begin{equation}
\widetilde{O}\Big(\kappa\gamma^{-1}\epsilon^{-m+1}\big(C_{\rm init}+\epsilon^{-1}(a+C_A)\big)\Big).
\end{equation}
\end{theorem}

The proof is essentially the same as Theorem \ref{QEE_thm}, the only difference is the number of sampling points, which has a quadratic saving, as a consequence of estimating eigenvalues on the one-dimensional real axis instead of the two-dimensional unit disk.

\begin{proof}
At step $l$,
\begin{equation}
N^{(l)} = \left\lceil \frac{2R^{(l)}}{\delta^{(l)}} \right\rceil = O(\kappa2^{(m-1)l}).
\end{equation}
Then the total sampling number $N_s$ is
\begin{equation}
\sum_{l=1}^{L} N^{(l)} = \kappa\sum_{l=1}^{L} O(2^{(m-1)l}) = O(\kappa\epsilon^{-(m-1)}).
\end{equation}
\end{proof}

Note that if $\mu = \lambda$ is an eigenvalue of $A$, then the eigenvector of $A$ associated with eigenvalue $\lambda$ is the right singular vector of $A-\lambda I$ associated with singular value 0, which is the ground state of $H_{\lambda}$. Running Algorithm \ref{QEE} gives an estimate $\tilde{\lambda}$ of an eigenvalue $\lambda$ of $A$, then the near-optimal ground state preparation algorithm from Ref. \cite{lin2020near} applied to Hamiltonian $H_{\tilde{\lambda}}$ produces an approximate eigenvector of $A$ associated with the eigenvalue $\lambda$.

\begin{theorem}[Quantum eigenvector preparation]\label{QEigenPrep}
Suppose that a matrix $A \in \mathbb{C}^{N \times N}$ is given through a $(1,a,0)$-block-encoding $U_A$ and an $\epsilon_0$-estimate of the eigenvalue $\lambda_0$, $\tilde{\lambda}_0$, is obtained. Then there is an algorithm that prepares the eigenvector associated with $\lambda_0$ to fidelity $1-\epsilon$ with probability $1-p_{\rm fail}$ using $O(a+\log(1/\gamma))$ ancilla qubits with gate count
\begin{equation}
\widetilde{O}\Big(\gamma^{-1}\big(C_{\rm init}+\Delta^{-1}(a+C_A)\big)\Big).
\end{equation}
\end{theorem}

\begin{proof}
For simplicity we treat $H_\mu$ as
\begin{equation}
\sqrt{(A-\mu I)^\dag(A-\mu I)}
\end{equation}
instead of the definition in Eq. \eqref{H_mod}, and the computed $H_\mu$ for the estimate $\tilde{\lambda}_0$ is referred to as
\begin{equation}
\widetilde{H}_{\lambda_0}= \sqrt{(A-\tilde{\lambda}_0I)^\dag(A-\tilde{\lambda}_0I)}.
\end{equation}
The spectrum of $H_{\lambda_0}$ is $\{ |\lambda_j-\lambda_0|, j=1,\dots,N \}$. The ground state preparation algorithm from Ref. \cite{lin2020near} assume a spectrum gap $\Delta$ between the ground state and the first excited state, hence one can guarantee the ground energy of $\widetilde{H}_{\lambda_0}$ is $|\lambda_0-\tilde{\lambda}_0|$ if $\epsilon_0 < \Delta/2$, and the ground states for $H_{\lambda_0}$ and $\widetilde{H}_{\lambda_0}$ are the same, which is denoted as $|E_0\rangle$. The ground state preparation algorithm output a state $|\tilde{\lambda}\rangle$ such that $|\langle E_0|\tilde{\lambda}\rangle| > 1-\epsilon$. The computational cost simply is the resource needed by the ground state preparation algorithm, and note that a query of $H_\mu$ requires $O(\log(1/\epsilon_0)) = O(\log(1/\Delta))$ queries of $U_A$ and $O((2a+4)\log(1/\Delta))$ primitive gates.
\end{proof}

For large-scale matrices, it is unreasonable to estimate all of the eigenvalues and often more desired to estimate the extreme eigenvalues. The sampling scheme in Algorithm \ref{QEE} can be modified to accomplish this task. Here we are solely concerned with the eigenvalue with the smallest modulus. The largest-modulus eigenvalue can be obtained similarly by considering the matrix $A^{-1}$.

\begin{theorem}[Extreme eigenvalue]\label{extreme}
Suppose that matrix $A \in \mathbb{C}^{N \times N}$ is given through a $(1,a,0)$-block-encoding $U_A$ and 0 is not an eigenvalue of $A$ (that is, nonsingular). Then there is an algorithm that, with probability at least $1-p_{\rm fail}$, outputs an $\epsilon$-estimate to the eigenvalue with the smallest modulus using $O(a+\log(1/\gamma))$ ancilla qubits with gate count
\begin{equation}
\widetilde{O}\Big(\kappa^2\gamma^{-1}\epsilon^{-2m+1}\big(C_{\rm init}+\epsilon^{-1}(a+C_A)\big)\Big).
\end{equation}
\end{theorem}

\begin{figure}
  \centering
  \includegraphics[scale=0.5]{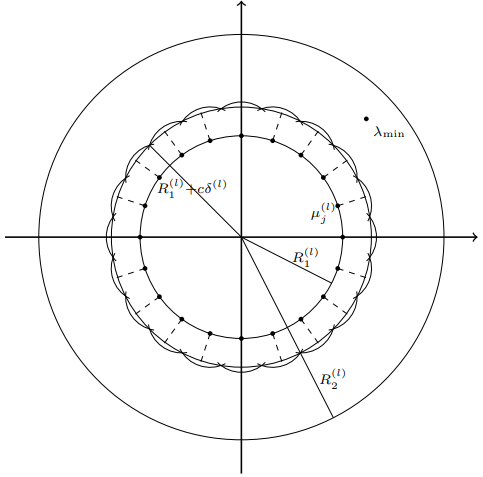}
  \caption{Illustration of the Case 2 in the proof of Theorem \ref{extreme}. $\lambda_{\min}$ lies outside all the circles centered at $\mu^{(l)}_j$ with radius $\delta^{(l)}$ (only partial arcs of these circles are shown in the diagram). Particularly, $\lambda_{\min}$ is outside the circle centered at the origin with radius $R_1^{(l)}+c\delta^{(l)}$. }\label{extreme_fig}
\end{figure}

\begin{proof}
We only describe the sampling scheme for locating the eigenvalue with the smallest modulus. The computational cost can be calculated similar to Theorem \ref{QEE}.

Take $\mu = 0$ in Eq. \eqref{defective}, we have that
\begin{equation}
\sigma_0(0) \le \min\limits_j |\lambda_j| \le 3(\kappa\sigma_0(0))^{1/m}.
\end{equation}
That is, the smallest-modulus eigenvalue $\lambda_{\min}$ lies in the annulus region $\{z\colon R_1^{(1)}=\sigma_0(0) \le |z| \le R_2^{(1)}=\min\{ 3(\kappa\sigma_0(0))^{1/m}, 1 \} \}$.
At step $l \ge 1$, we sample $M$ points $\{ \mu^{(l)}_j = R_1^{(l)}e^{ij2\pi/M}, 0\le j\le M-1 \}$ on the circle $\{ z\colon |z|=R_1^{(l)} \}$. Let $\delta^{(l)} = \frac{1}{\kappa}\Big( \frac{1}{3}\frac{R_2^{(l)}-R_1^{(l)}}{2} \Big)^m$. There are two cases:
\begin{enumerate}[1)]
  \item Case 1: there exists some $j_0$ such that $\sigma_0(\mu^{(l)}_{j_0}) \le \delta^{(l)}$, then according to Eq. \eqref{defective}, we can update $R_1^{(l)}$ and $R_2^{(l)}$ to
\begin{equation}
\left\{
\begin{aligned}
& R_1^{(l+1)} = R_1^{(l)}, \\
& R_2^{(l+1)} = R_1^{(l)}+3(\kappa\sigma_0(\mu^{(l)}_{j_0}))^{1/m} \le R_1^{(l)}+3(\kappa \delta^{(l)})^{1/m} \le R_1^{(l)}+\frac{R_2^{(l)}-R_1^{(l)}}{2} = \frac{R_1^{(l)}+R_2^{(l)}}{2}.
\end{aligned}
\right.
\end{equation}
  \item Case 2: $\sigma_0(\mu^{(l)}_j) > \delta^{(l)}$ for all $j$, then $\lambda_{\min}$ is outside all the circles centered at $\mu^{(l)}_j$ with radius $\delta^{(l)}$ (Fig. \ref{extreme_fig}). By taking $M^{(l)} = \big\lceil \pi/(\arcsin(\sqrt{1-c^2}\delta^{(l)}/R_1^{(l)})) \big\rceil$ for some constant $c < 1$ sufficiently close to 1, the circle centered at the origin with radius $R_1^{(l)}+c\delta^{(l)}$ is covered by these circles. Hence we can replace $R_1^{(l)}$ by $\widetilde{R}_1^{(l)} = R_1^{(l)}+c\delta^{(l)}$. Keep replacing $\widetilde{R}_1^{(l)}$ as above until Case 1 happens. Then we can update $R_1^{(l)}$ and $R_2^{(l)}$ to
\begin{equation}
\left\{
\begin{aligned}
& R_1^{(l+1)} = \widetilde{R}_1^{(l)}, \\
& R_2^{(l+1)} = R_2^{(l)}.
\end{aligned}
\right.
\end{equation}
\end{enumerate}
At some step $L$, Case 1 happens with the updated $R_1^{(L+1)}$ and $R_2^{(L+1)}$ satisfying $R_2^{(L+1)}-R_1^{(L+1)} < \epsilon$, then $\mu^{(L)}_{j_0}$ serves as an estimate with additive error $\epsilon$ for the extreme eigenvalue.

Note that a step of Case 2 must be followed by one or more steps of Case 1. Since Case 1 shrinks the annulus by a factor of 2, there are at most $O(\log(1/\epsilon))$ Case 1 steps. In each Case 2 step, one replace $R_1^{(l)}$ at most $O\big((R_2^{(l)}-R_1^{(l)})/\delta^{(l)}\big) = O(\kappa\epsilon^{-m+1})$ times. Therefore, the total sampling number is
\begin{equation}
\begin{aligned}
\sum_{l=1}^{\lceil\log(1/\epsilon)\rceil} O(\kappa\epsilon^{-m+1} \cdot M^{(l)}) &= O(\kappa\epsilon^{-m+1} \cdot \kappa\epsilon^{-m}) \\
&= O(\kappa^2\epsilon^{-2m+1}).
\end{aligned}
\end{equation}
\end{proof}

\begin{figure}
  \centering
  \includegraphics[scale=0.5]{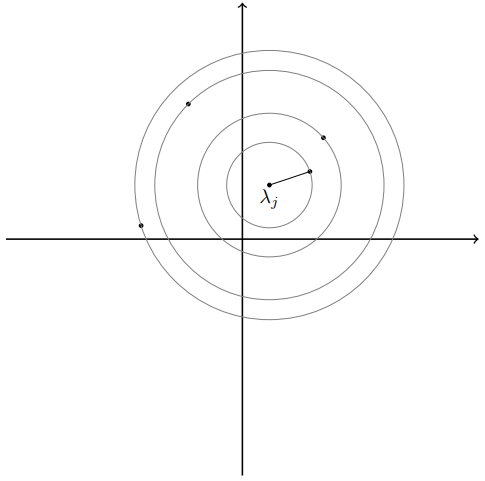}
  \caption{Diagram of the point gap. The solid dots represent the eigenvalues of $A$. The radius of the innermost circle is the point gap at $\lambda_j$. One can successively estimate the eigenvalues closest to $\lambda_j$ by using the method of Theorem \ref{extreme}, proceeding from the closest to the next in order.}\label{point_gap_fig}
\end{figure}

Note that for an eigenvalue $\lambda_j$ of $A$, the modulus of the non-zero smallest-modulus eigenvalue of $A-\lambda_jI$ is the spectral gap of $A$ at $\lambda_j$ (Fig. \ref{point_gap_fig}). Hence, with the knowledge of an eigenvalue $\lambda_j$, one can approximate the spectral gap at $\lambda_j$ by running the extreme eigenvalue algorithm for the matrix $A-\lambda_jI$ and evaluating the modulus of the output.

Similar to the point gap problem discussed above is the line gap problem. Given a reference line $\bf{L}$, the goal now is to output an eigenvalue $\lambda_{\min}$ closest to $\bf{L}$ up to an accuracy $\epsilon$. For simplicity we take $\bf{L}$ to be the real axis and assume that ${\rm Im}(\lambda_j)>0$ without loss of generality.

\begin{theorem}[Line gap problem]\label{line_gap}
Suppose that matrix $A \in \mathbb{C}^{N \times N}$ is given through a $(1,a,0)$-block-encoding $U_A$ with ${\rm Im}(\lambda_j)>0$. Then there is an algorithm that, with probability at least $1-p_{\rm fail}$, outputs an $\epsilon$-estimate to the eigenvalue closest to the real axis using $O(a+\log(1/\gamma))$ ancilla qubits with gate count
\begin{equation}
\widetilde{O}\Big(\kappa^2\gamma^{-1}\epsilon^{-2m+1}\big(C_{\rm init}+\epsilon^{-1}(a+C_A)\big)\Big).
\end{equation}
\end{theorem}
\begin{proof}
See Appendix \ref{append_line_gap}.
\end{proof}

\section{Applications}\label{app}

As discussed in \cite{3n8f-k8pl}, non-Hermitian eigenproblems arise across diverse fields, from estimating Liouvillian gaps and detecting spontaneous PT-symmetry breaking to analyzing classical Markov processes. Here We present two additional applications: analyzing the stability of dynamical systems and solving polynomial root-finding problems.

Consider a dynamical system governed by the differential equation $dy/dt = F(y)$ where $y$ is some vector-valued function of $t$ and $F$ is a function from $\mathbb{R}^N$ to itself. Stability is often studied in terms of the steady state solution, which is, by definition, the limit of $y(t)$ as $t$ tends to infinity. In most cases the stability of a dynamical system can be determined by its linear stability, i.e., by the stability of the linear approximation of $F$ at $\bar{y}$. Specifically, the system is stable if all the eigenvalues of the Jacobian matrix
\begin{equation}
J = \bigg\{ \frac{\partial f_i(\bar{y})}{\partial x_j} \bigg\}_{i,j = 1,\dots,N}
\end{equation}
have negative real parts and unstable if at least one eigenvalue has a positive real part \cite{saad2011numerical}. One can distinguish the two cases by determining the existence of an eigenvalue on the right half of the unit disk, which can be accomplished by Algorithm \ref{QEE} (One can modify the algorithm so that it samples points on the right half of the unit disk and returns ``no eigenvalue" if line 14 in Algorithm \ref{QEE} is not satisfied for all sampled points). If some eigenvalues of $J$ lie on the imaginary axis, then the stability of the system cannot be determined by its linear stability \cite{guckenheimer2013nonlinear}. This situation can be identified by employing the eigenvalue estimation algorithm on the imaginary axis, similar to Theorem \ref{QEE_real}. Note that Jacobian matrices typically are large nonsymmetric and sparse matrices, for example when $F$ arises from discretizing a partial differential operator. 

%
Another application arises from the fact that eigenvalue evaluation is equivalent to polynomial root-finding. Through the characteristic polynomial of the matrix, eigenvalue problems can be reduced to polynomial root-finding problems. Conversely, any polynomial root-finding problem can be stated as an eigenvalue problem via the companion matrix of the polynomial, which is non-normal and sparse \cite{trefethen2022numerical}. Although determining the roots of a polynomial is a notoriously ill-conditioned problem in general \cite{trefethen2022numerical}, there are well-conditioned cases. Algorithm \ref{QEE} applied to well-conditioned polynomials would output one of its roots, and in particular, Theorem \ref{extreme} gives a root with the largest or smallest modulus.

\section{Numerical Simulation}\label{numer}

To validate the correctness of the classical-quantum hybrid framework and the iterative refinement logic of our algorithm, we performed numerical simulations. As these are hybrid algorithms whose key quantum subroutines—such as constructing 
 $H_\mu$ via the Quantum Singular Value Transformation (QSVT) and performing near-optimal ground state energy estimation—are computationally prohibitive to simulate classically, our numerical verification focuses on the classical component.  Specifically, we bypass the quantum ground state estimation by directly computing $\sigma_0(A-\mu I)$ using MATLAB functions, rather than explicitly building and diagonalizing the Hermitian matrix $\sqrt{(A-\mu I)^\dag(A-\mu I)}$. We test Algorithm \ref{QEE} and the extreme eigenvalue algorithm (Theorem \ref{extreme}) for matrices with known eigenvalues, and the algorithms output results with prescribed accuracy \cite{quantumeigensolver}. We present the experimental result in Table \ref{experiment}. Note that Algorithm \ref{QEE} can only output one of the eigenvalues of the input matrix, and one may adjust the sampling scheme to obtain eigenvalues from a certain region.

\begin{table}
  \centering
  \caption{The experimental result of four matrix instances with precision $\epsilon=0.01$. The second column are the spectrum of the matrix instances and the superscripts indicate the multiplicity of the corresponding eigenvalue. The third column is the output estimate of Algorithm \ref{QEE} and the fourth column is the estimated extreme eigenvalue.}\label{experiment}
  \begin{tabular}{c|c|c|c}
  \hline
  Instance & Eigenvalues & Output estimate & Estimated extreme eigenvalue \\
  \hline
  dim=4,\,m=1 & 0.25+0.25i,\,1$^{(3)}$ & 0.2474+0.2500i & 0.2519+0.2480i \\
  dim=4,\,m=1 & 0.125+0.125i,\,$-$0.25$-$0.25i,\,1$^{(2)}$ & $-$0.25$-$0.25i & 0.1261+0.1239i \\
  dim=4,\,m=2 & 0.3904$^{(2)}$,\,0.7808$^{(2)}$ & 0.3888$-$0.0001i & 0.3893 \\
  dim=8,\,m=2 & 0.25+0.25i$^{(2)}$,\,$-$0.5$^{(2)}$,\,0.5i$^{(2)}$,\,0.75,\,1 & $-$0.5013 & 0.2493+0.2493i \\
  \hline
  \end{tabular}
\end{table}

\section{Discussion}\label{dis}

The query complexity of $U_A$ is dominated by the accuracy parameter $\epsilon$, which is $O(\epsilon^{-2m+1})$. Hence our algorithm is feasible for $m=O(1)$. Also, we require $\kappa = O({\rm poly}(n))$ so that the grid spacing $\delta^{(l)} = 1/(\kappa(3\cdot2^l)^m)$ is feasible. More importantly, this is the regime that eigenvalue analysis is meaningful. Matrices with large condition number is in some sense far from normal, for which eigenvalue analysis may fail, and it is the concept of pseudospectra that plays a crucial role in analysing these matrices \cite{trefethen2020spectra}. On the other hand, establishing the lower bound on the query complexity for this problem would be a promising direction for future research.

As depicted in Fig. \ref{point_gap_fig}, one can estimate the eigenvalues iteratively, starting from an estimated eigenvalue $\lambda_j$ and then identifying eigenvalues in order of their proximity to $\lambda_j$ using the method of Theorem \ref{extreme}. It is of interest to explore if there are more efficient ways of estimating multiple eigenvalues of a non-normal matrix.

In conclusion, we have presented a quantum algorithm that significantly improves the complexity for a fundamental linear algebra problem. The central insight of our work—the reduction of non-normal eigenvalue estimation to ground state energy estimation—is likely to have lasting value. While direct experimental demonstration on large-scale problems awaits fault-tolerant quantum hardware, the theoretical superiority of our approach is unequivocal. Our algorithm not only achieves a proven polynomial speedup but does so by leveraging the particularly robust and rapidly advancing subroutine in quantum computation, making it the most promising pathway towards solving non-normal eigenvalue problems on quantum computers.

\section{Acknowledgement}
This work was supported by the National Natural Science Foundation special project of China (Grant No. 12341103). The National Key R\&D Program of China (Grant No. 2023YFA1009403) and National Natural Science Foundation of China (Grant No. 62372444).

\bibliographystyle{unsrt}
\bibliography{Ref}

\appendix

\section{\label{proof_lemma_1}Proof of Lemma \ref{lemma}}

We follow the proof of \cite{3n8f-k8pl} with some modifications.
\begin{proof}[Proof of Lemma \ref{lemma}]
Inequalities \eqref{diag} and \eqref{defective} are trivial if $\mu$ is an eigenvalue of $A$. Suppose that $\mu$ is not an eigenvalue of $A$, then $A-\mu I$ is non-singular. For operator norm, $\|A\| = \sigma_{\rm max}(A)$. Hence,
\begin{equation}
\begin{aligned}
\sigma_0(\mu) &= \sigma_{\rm min}(A-\mu I) = \frac{1}{\sigma_{\rm max}[(A-\mu I)^{-1}]} \\
&= \| (A-\mu I)^{-1} \|^{-1}.
\end{aligned}
\end{equation}
Since $\|A^{-1}\|^{-1} \le |\lambda_j| \le \|A\|$ for non-singular $A$, we have
\begin{equation}
\sigma_0(\mu) = \| (A-\mu I)^{-1} \|^{-1} \le |\lambda_j-\mu|.
\end{equation}
Taking the minimum over all eigenvalues, we obtain
\begin{equation}\label{lb}
\sigma_0(\mu) \le \min\limits_j |\lambda_j-\mu|.
\end{equation}

On the other hand,
\begin{equation}\label{RHS}
\begin{aligned}
\| (A-\mu I)^{-1} \|^{-1} &= \| P(J-\mu I)^{-1}P^{-1}\|^{-1} \\
&\ge \kappa^{-1} \|(J-\mu I)^{-1}\|^{-1}.
\end{aligned}
\end{equation}
If $A$ is diagonalizable, then
\begin{equation}\label{RHS_diag}
\|(J-\mu I)^{-1}\|^{-1} = \min\limits_j |\mu - \lambda_j|,
\end{equation}
and Eq. \eqref{diag} follows. If $A$ is defective, then $\|(J - \mu I)^{-1}\|^{-1}$ is the smallest singular value among all the Jordan matrices. Let $m_j$ be the block size of the largest Jordan block in Jordan matrix $J(\lambda_j)$. According to Ref. \cite{kahan1982residual}, the smallest singular value for a typical Jordan matrix is lower bounded by $\frac{|\lambda_j-\mu|^{m_j}}{(1+|\lambda_j-\mu|)^{m_j-1}}$. Therefore,
\begin{equation}\label{SSV_Jordan}
\|(J-\mu I)^{-1}\|^{-1} \ge \min\limits_j\frac{|\lambda_j-\mu|^{m_j}}{(1+|\lambda_j-\mu|)^{m_j-1}}.
\end{equation}
Suppose that the minimum of Eq. \eqref{SSV_Jordan} is achieved at $\lambda_{j_0}$.
Note that $|\lambda_j-\mu| \le 2$, we have
\begin{equation}\label{scaling}
\begin{aligned}
&\frac{|\lambda_{j_0}-\mu|^{m_{j_0}}}{(1+|\lambda_{j_0}-\mu|)^{m_{j_0}-1}} \\ = &\left(\frac{|\lambda_{j_0}-\mu|}{1+|\lambda_{j_0}-\mu|}\right)^{m_{j_0}}(1+|\lambda_{j_0}-\mu|) \\ \ge &\left(\frac{|\lambda_{j_0}-\mu|}{3}\right)^{m_{j_0}}.
\end{aligned}
\end{equation}
Combining \eqref{RHS} with \eqref{scaling} gives
\begin{equation}
\sigma_0(\mu) \ge \kappa^{-1} \left(\frac{|\lambda_{j_0}-\mu|}{3}\right)^{m_{j_0}},
\end{equation}
that is, 
\begin{equation}\label{upper_bound_for_j_0}
|\lambda_{j_0}-\mu| \le 3(\kappa \sigma_0(\mu))^{1/m_{j_0}}.
\end{equation}
Therefore,
\begin{equation}\label{RHS_defective}
\min\limits_j |\lambda_j-\mu| \le 3(\kappa \sigma_0(\mu)))^{1/m}
\end{equation}
for some $m = m_{j_0} \le \max\limits_j m_j$.
\end{proof}

\section{Algorithm \ref{QEE} from the perspective of pseudospectra}\label{pseudospectra}

Algorithm \ref{QEE} can be interpreted using the concept of pseudospectra. Let $\Lambda(A)$ denote the spectrum of $A$. There are several equivalent ways of defining the pseudospectra of $A$ \cite{trefethen2020spectra}. Here we adopt the definition in terms of the resolvent of $A$ at $\mu$, $(\mu I-A)^{-1}$, and for $\|\cdot\| = \|\cdot\|_2$ we have the following definition.

\begin{definition}
Let $A \in \mathbb{C}^{N \times N}$ and $\epsilon > 0$ be arbitrary. The $\epsilon$-pseudospectrum $\Lambda_\epsilon(A)$ of $A$ is the set of $\mu \in \mathbb{C}$ such that
\begin{equation}
\sigma_0(\mu) = \|(A-\mu I)^{-1}\|^{-1} \le \epsilon.
\end{equation}
\end{definition}

From the definition, it follows that the pseudospectra of $A$ are nested sets,
\begin{equation}
\Lambda_{\epsilon_1} \subseteq \Lambda_{\epsilon_2}, 0 < \epsilon_1 \le \epsilon_2,
\end{equation}
and that the intersection of all the pseudospectra is the spectrum,
\begin{equation}
\underset{\epsilon > 0}{\bigcap}\Lambda_\epsilon(A) = \lim\limits_{\epsilon\to0}\Lambda_\epsilon(A) = \Lambda(A).
\end{equation}

By definition, the condition $\sigma_0(\mu^{(l)}_{j,k}) \le \delta^{(l)}$ in line 14 of Algorithm \ref{QEE} indicates that $\mu^{(l)}_{j,k}$ belongs to the $\delta^{(l)}$-spectrum of $A$. As $l \to \infty$, $\delta^{(l)} \to 0$. Hence $\mu^{(l)}$ converges to an eigenvalue in the limit of $l \to \infty$. 

At the end of this section we reformulate Lemma \ref{lemma} in terms of $\Lambda_\epsilon(A)$, making it a direct generalization of the Bauer-Fike Theorem \cite[Theorem 2.3]{trefethen2020spectra}. We use the notation $\Delta_\epsilon$ for an $\epsilon$-ball $\Delta_\epsilon = \{ z\in\mathbb{C} \colon |z| \le \epsilon \}$. A sum of sets means
\begin{equation}
\Lambda(A)+\Delta_\epsilon = \{ z\colon z=z_1+z_2, z_1 \in \Lambda(A), z_2\in \Delta_\epsilon \}.
\end{equation}

\begin{lemma}
Let $A = PJP^{-1}$ be the Jordan decomposition of $A$. Suppose that $\|A\| \le 1$ and the Jordan condition number $\|P\|\|P^{-1}\| \le \kappa$. Let $m_{\rm max}$ denote the block size of the largest Jordan block. Then
\begin{equation}
\Lambda(A)+\Delta_\epsilon \subseteq \Lambda_\epsilon(A) \subseteq \Lambda(A)+\Delta_{\epsilon\kappa}
\end{equation}
if $A$ is diagonalizable, and
\begin{equation}
\Lambda(A)+\Delta_\epsilon \subseteq \Lambda_\epsilon(A) \subseteq \Lambda(A)+\Delta_{3(\epsilon\kappa)^{1/m}}
\end{equation}
for some $m \le m_{\rm max}$ if $A$ is defective.
\end{lemma}
\begin{proof}
For $\mu \in \Lambda(A)+\Delta_\epsilon$, $\min_j|\mu-\lambda_j| \le \epsilon$, so according to \eqref{lb}
\begin{equation}
\sigma_0(\mu) \le \min\limits_j|\mu-\lambda_j| \le \epsilon,
\end{equation}
that is, $\mu \in \Lambda_\epsilon(A)$, implying that $\Lambda(A)+\Delta_\epsilon \subseteq \Lambda_\epsilon(A)$.

For $\mu \in \Lambda_\epsilon(A)$, equations \eqref{RHS} and \eqref{RHS_diag} imply that
\begin{equation}
\kappa^{-1} \cdot \min\limits_j |\mu-\lambda_j| \le \sigma_0(\mu) \le \epsilon 
\end{equation}
for diagonalizable matrices. Hence $\mu \in \Lambda(A)+\Delta_{\epsilon\kappa}$. Similarly, for defective matrices, from Eq. \eqref{RHS_defective} we have
\begin{equation}
\min\limits_j |\lambda_j-\mu| \le 3(\kappa \sigma_0(\mu)))^{1/m} \le 3(\epsilon\kappa))^{1/m},
\end{equation}
which shows that $\mu \in \Lambda(A)+\Delta_{3(\epsilon\kappa)^{1/m}}$.
\end{proof}

\section{Proof of Lemma \ref{H_mu}}\label{proof_lemma_2}

\begin{figure}[htbp]
\centering
\includegraphics[scale=0.4]{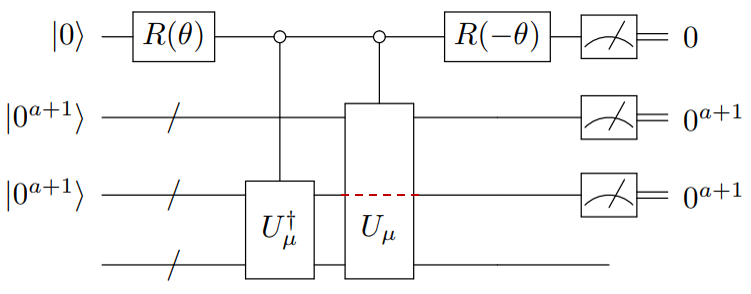}
\caption{\label{BE_circuit}The block encoding circuit of $(A-\mu I)^{\dagger}(A-\mu I)/\alpha_\mu^2+\nu I$ with normalization factor $1+\nu$. $R(\theta) = \begin{pmatrix} \cos\theta & -\sin\theta \\ \sin\theta & \cos\theta\end{pmatrix}$ and $\theta = \arccos(\sqrt{1/(1+\nu)})$. The dashed line indicates that the unitary $U_\mu$ does not act on these qubits.}
\end{figure}

\begin{proof}[Proof of Lemma \ref{H_mu}]

We construct $H_\mu$ by quantum eigenvalue transformation using quantum signal processing (QSP) technique.

\begin{lemma}[QSP for polynomials of arbitrary parity \cite{gilyen2019quantum}]\label{QSP}
Let $U$ be an $(\alpha,a,0)$-block-encoding of a Hermitian matrix $H$. Let $P \in \mathbb{R}[x]$ be a degree-$d$ real polynomial and $|P(x)| \le 1$ for $x \in [-1,1]$. Then there exists a $(1,a+2,0)$-block-encoding $\tilde{U}$ of $P(H/\alpha)$ using $O(d)$ queries of $U$ or $U^\dag$, and $O((a+1)d)$ primitives quantum gates.
\end{lemma}

To this end, we show that the square function restricted to $[\eta,1]$ can be approximated by a polynomial satisfying the conditions of Lemma \ref{QSP}.

\begin{lemma}[Polynomial approximation of the square function]\label{square}
For $\epsilon \in (0,1)$, there exists a polynomial $p \in \mathbb{R}[x]$ of degree $O(\log(1/\epsilon))$ such that
\begin{enumerate}[1)]
  \item $|p(x)| \le 1, x \in [-1,1]$;
  \item $|p(x) - \sqrt{x}| \le \epsilon, x \in [\eta, 1]$.
\end{enumerate}
\end{lemma}

\begin{proof}[Proof of Lemma \ref{square}]

We first approximate the square function on the interval $[\eta/2,1]$ by a $d$-degree polynomial $p_1(x)$. Let $L(y)$ be the linear function that maps $[-1,1]$ to $[-\eta,1]$. Then
\begin{equation}
f(y) = \sqrt{L(y)} \colon [-1,1] \to \mathbb{R}
\end{equation}
is Lipschitz continuous on $[-1,1]$, it has a unique representation as a Chebyshev series \cite[Theorem 3.1]{trefethen2019approximation}
\begin{equation}
f(y) = \sum_{k=0}^{\infty} a_kT_k(y).
\end{equation}
Consider the polynomial by truncation of the series to degree $d$,
\begin{equation}
f_d(y) = \sum_{k=0}^{d} a_kT_k(y).
\end{equation}
According to the Bernstein Theorem for Chebyshev coefficients of a analytic function \cite[Theorem 8.1]{trefethen2019approximation}, for $f(y)$ considered here, its Cheyshev coefficients $a_k$ satisfy
\begin{equation}
\begin{aligned}
& |a_0| \le 1; \\
& |a_k| \le 2\rho^{-k} \text{ for some } \rho > 1, k \ge 1.
\end{aligned}
\end{equation}
We have
\begin{equation}
\begin{aligned}
|f(y)-f_d(y)| &= \left|\sum_{k=d+1}^{\infty} a_kT_k(y)\right| \\
&\le 2\sum_{k=d+1}^{\infty} \rho^{-k} \\
&= \frac{2}{\rho^d(\rho-1)}.
\end{aligned}
\end{equation}
Hence, for the truncation error to be smaller than some $\epsilon_1$, we need $d = O(\log(1/\epsilon_1))$, and
\begin{equation}
p_1(x) = f_d(L^{-1}(x))
\end{equation}
is the polynomial that approximate the square function on $[\eta/2,1]$ to precision $\epsilon_1$. Due to the aliasing of Chebyshev polynomials \cite[Theorem 4.2]{trefethen2019approximation}, such a polynomial can be explicitly constructed by Lagrange interpolation through Chebyshev points scaled to the interval $[\eta/2,1]$, and the interpolation error bound is only different from the truncation error bound by a factor of 2 \cite[equation (4.9)]{trefethen2019approximation}.

For the polynomial to be bounded by 1 on $[-1,1]$, we consider the shifted Heaviside function
\begin{equation}
H(x-3\eta/4) = \left\{
\begin{aligned}
& 0, x < 3\eta/4, \\
& 1, x \ge 3\eta/4.
\end{aligned}
\right.
\end{equation}
It is well known that $H(x-3\eta/4)$ can be approximated to $\epsilon_2$ precision by an polynomial of degree $O(\frac{1}{\eta}\log\frac{1}{\epsilon_2})$ \cite[Corollary 7]{low2017hamiltonian}, such that
\begin{equation}
\begin{aligned}
& |p_2(x)-H(x-3\eta/4)| \le \epsilon_2, x \in [-1,\eta/2] \cup [\eta,1]; \\
& |p_2(x)| \le 1, x \in [-1,1].
\end{aligned}
\end{equation}
Let $p(x) = p_1(x)p_2(x)$. Then $|p(x)| \le 1$ for $x \in [-1,1]$ and
\begin{equation}
\begin{aligned}
|\sqrt{x} - p(x)| 
&\le |\sqrt{x} - p_1(x)| + |p_1(x)-p_1(x)p_2(x)| \\
&\le \epsilon_1 + (1+\epsilon_1)\epsilon_2 \\
&\le \epsilon
\end{aligned}
\end{equation}
for $x \in [\eta,1]$ by setting $\epsilon_1 = \frac{\epsilon}{2}$ and $\epsilon_2 = \frac{\epsilon}{2(1+\epsilon_1)}$.

The degree of $p(x)$ is
\begin{equation}
O\left(\log\frac{1}{\epsilon_1}\right)+O\left(\frac{1}{\eta}\log\frac{1}{\epsilon_2}\right) = O\left(\log\frac{1}{\epsilon}\right).
\end{equation}
Here we treat $\eta$ as a constant and hide it in the big-$O$ notation.
\end{proof}
Since the block encoding of $(A-\mu I)^{\dagger}(A-\mu I)/\alpha_\mu^2+\nu I$ requires $2a+3$ ancilla qubits (Fig. \ref{BE_circuit}), Lemma \ref{H_mu} follows by applying Lemma \ref{QSP} and Lemma \ref{square} to
\begin{equation}
(A-\mu I)^{\dagger}(A-\mu I)/\alpha_\mu^2+\nu I.
\end{equation}
\end{proof}

\section{Proof of Theorem \ref{line_gap}}\label{append_line_gap}

\begin{figure}
  \centering
  \includegraphics[scale=0.5]{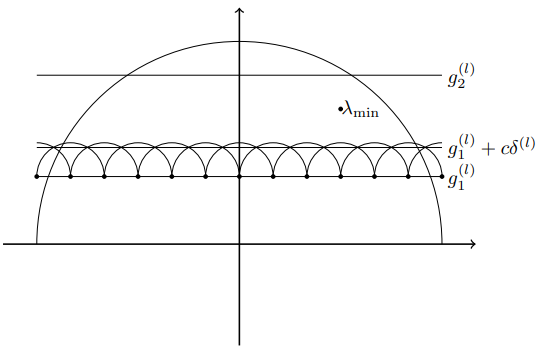}
  \caption{Illustration of the Case 2 in the proof of Theorem \ref{line_gap}. The solid points are the sampled values of $\mu^{(l)}_j$. $\lambda_{\min}$ lies outside all the circles centered at $\mu^{(l)}_j$ with radius $\delta^{(l)}$ (only partial arcs of these circles are shown in the diagram). Particularly, ${\rm Im}(\lambda_{\rm min}) > g_1^{(l)}+c\delta^{(l)}$.}\label{line_gap_fig}
\end{figure}

\begin{proof}
Similar to the proof of Theorem \ref{extreme}, we only describe the sampling scheme for locating the eigenvalue closest to the real axis.

We initialize the sample region as $\{ z\colon g_1^{(1)} = 0 \le {\rm Im}(z) \le g_2^{(1)} = 1 \}$. At step $l \ge 1$, set $\delta^{(l)} = \frac{1}{\kappa}\Big( \frac{1}{3}\frac{g_2^{(l)}-g_1^{(l)}}{2} \Big)^m$. We sample $2M-1$ points $\{ \mu^{(l)}_j = \pm j\delta^{(l)}+g_1^{(l)}i, 0\le j\le M-1 \}$ on the line $\{ z\colon {\rm Im}(z)=g_1^{(l)} \}$. There are two cases:
\begin{enumerate}[1)]
  \item Case 1: there exists some $j_0$ such that $\sigma_0(\mu^{(l)}_{j_0}) \le \delta^{(l)}$, then according to Eq. \eqref{defective}, we can update $g_1^{(l)}$ and $g_2^{(l)}$ to
\begin{equation}
\left\{
\begin{aligned}
& g_1^{(l+1)} = g_1^{(l)}, \\
& g_2^{(l+1)} = g_1^{(l)}+3(\kappa\sigma_0(\mu^{(l)}_{j_0}))^{1/m} \le g_1^{(l)}+3(\kappa \delta^{(l)})^{1/m} \le g_1^{(l)}+\frac{g_2^{(l)}-g_1^{(l)}}{2} = \frac{g_1^{(l)}+g_2^{(l)}}{2}.
\end{aligned}
\right.
\end{equation}
  \item Case 2: $\sigma_0(\mu^{(l)}_j) > \delta^{(l)}$ for all $j$, then $\lambda_{\min}$ is outside all the circles centered at $\mu^{(l)}_j$ with radius $\delta^{(l)}$ (Fig. \ref{line_gap_fig}). By taking $M^{(l)} = \big\lceil 1/(2\sqrt{(1-c^2)}\delta^{(l)}) \big\rceil$ for some constant $c < 1$ sufficiently close to 1, the rectangular region $\{ z\colon g_1^{(l)}\le{\rm Im}(z)\le g_1^{(l)}+c\delta^{(l)} \}$ is covered by these circles. Hence we can replace $g_1^{(l)}$ by $\tilde{g}_1^{(l)} = g_1^{(l)}+c\delta^{(l)}$. Keep replacing $\tilde{g}_1^{(l)}$ as above until Case 1 happens. Then we can update $g_1^{(l)}$ and $g_2^{(l)}$ to
\begin{equation}
\left\{
\begin{aligned}
& g_1^{(l+1)} = \tilde{g}_1^{(l)}, \\
& g_2^{(l+1)} = g_2^{(l)}.
\end{aligned}
\right.
\end{equation}
\end{enumerate}
At some step $L$, Case 1 happens with the updated $g_1^{(L+1)}$ and $g_2^{(L+1)}$ satisfying $g_2^{(L+1)}-g_1^{(L+1)} < \epsilon$, then $\mu^{(L)}_{j_0}$ serves as an estimate with additive error $\epsilon$ for the extreme eigenvalue.

Note that a step of Case 2 must be followed by one or more steps of Case 1. Since Case 1 shrinks the annulus by a factor of 2, there are at most $O(\log(1/\epsilon))$ Case 1 steps. In each Case 2 step, one replace $g_1^{(l)}$ at most $O\big((g_2^{(l)}-g_1^{(l)})/\delta^{(l)}\big) = O(\kappa\epsilon^{-m+1})$ times. Therefore, the total sampling number is
\begin{equation}
\begin{aligned}
\sum_{l=1}^{\lceil\log(1/\epsilon)\rceil} O(\kappa\epsilon^{-m+1} \cdot M^{(l)}) &= O(\kappa\epsilon^{-m+1} \cdot \kappa\epsilon^{-m}) \\
&= O(\kappa^2\epsilon^{-2m+1}).
\end{aligned}
\end{equation}
\end{proof}

\end{document}